\documentclass[aps,twocolumn,prd,showpacs,nofootinbib]{revtex4}
\usepackage{amsmath}
\usepackage{graphicx}
\usepackage{subfigure}
\usepackage{dcolumn}
\usepackage{bm}
\usepackage{amssymb}
\usepackage{latexsym}

\newcommand{\be}{\begin{equation}}
\newcommand{\ee}{\end{equation}}

\def\spose#1{\hbox to 0pt{#1\hss}}
\def\lta{\mathrel{\spose{\lower 3pt\hbox{$\mathchar"218$}}
     \raise 2.0pt\hbox{$\mathchar"13C$}}}
\def\gta{\mathrel{\spose{\lower 3pt\hbox{$\mathchar"218$}}
     \raise 2.0pt\hbox{$\mathchar"13E$}}}

\newcommand{\de}[2]{\kern - #1 em \mathrm{d} #2}

\begin{document}

\title{A Quantum Gravitational Relaxation of 
The Cosmological Constant}

\author{Stephon Alexander}
\email{stephon@itp.stanford.edu}
\affiliation{Stanford Linear Accelerator Center and ITP, Dept of Physics , Stanford University, 
2575 Sand Hill Rd., Menlo Park, CA 94025 USA}
\date{\today}

\begin{abstract}
Similar to QCD, general relativity has a $\Theta$ sector due to large diffeomorphisms.  We make explicit, for the first time, that the gravitational CP violating $\Theta$ parameter is non-perturbatively related  to the cosmological constant.     A gravitational pseudoscalar coupling to massive fermions gives rise to general relativity from a topological $B\wedge F$ theory through a chiral symmetry breaking mechanism.   We show that a gravitational Peccei-Quinn like mechanism can dynamically relax the cosmological constant.

\end{abstract}

\maketitle

\section{Introduction}

It has always been a dream to solve the cosmological constant problem by relaxing it to the minimum of a potential \cite{Weinberg:1988cp}.  This hope has been especially unsuccessful in conventional canonical quantum gravity.  In QCD, the strong CP problem was solved by relaxing the $\tilde{\Theta}$ parameter at the minimum of the potential associated with an axion field via the Peccei-Quinn mechanism.   It turns out that when quantum gravity is formulated in the Ashtekar-Sen variables (LQG), the theory has a semblance to Yang-Mills theory and the cosmological constant problem becomes analogous to the strong CP problem.  It is the purpose of this paper to make this analogy explicit and use a Peccei-Quinn like mechanism to pave a possible route to solving the cosmological constant problem. 

Loop Quantum Gravity (LQG) has a one parameter family of ambiguities which is labeled by $\gamma$, the Barbero-Immirizi parameter.  This parameter  plays a similar role to the QCD $\tilde{\Theta}$ parameter which labels the unitarily inequivalent sectors of the quantum theory.  These sectors can be accessed by tunneling events due to instanton field configurations.        $\tilde{\Theta}$ is also a measure of CP violation, which is constrained by the neutron electric dipole moment to be $\tilde{\Theta} \leq 10^{-9}$.   In LQG the Barbero-Immirizi parameter is also  a measure of CP violation, since it couples to the first Pontrjagin class.  Specifically, in the quantum theory $\gamma$ corresponds to unitarily inequivalent representations  of the algebra of geometric operators.   For example, the simplest eigenvalues of the area operator $\hat{A}_{s}$ in the $\gamma$ quantum sector is given by
\be A_{[j]} =8\pi\gamma l_{Pl}^{2}\sum_{I} \sqrt{j_{I}(j_{I} + 1)} \ee

What fixes the value of the Barbero-Immirizi parameter?  In this letter we will show that this question is connected to another parameter in general relativity, the cosmological constant.   Therefore, the question of fixing the Barbero-Immirizi parameter is related to a quantum gravitational determination of the cosmological constant.  Through this relationship we will propose a possible dynamical, background independent mechanism to relax the cosmological constant.

Specifically, we will demonstrate that a Peccei-Quinn-like mechanism associated with a non-vanishing vev of fermion bilinears coupled to gravity will yield an effective potential for a parameter which alters the ratio of Barbero-Immirizi parameter and the cosmological constant.   We then use the value of the Barbero-Immirizi parameter determined from Black Hole quasinormal modes to determine the    
conditions which relaxes the cosmological constant.  This mechanism is deeply tied to the presence of gravitational instantons in the state space of Loop Quantum Gravity which arise naturally as exact solutions of the (anti)self dual formulation of general relativity\cite{Soo1}.\footnote{While we do not perform any explicit instanton calculations in this paper, it was suggested by Soo \cite{Soo1} that, through a bundle reduction, instantons can reduce the value of cosmological constant.}

This paper is organized as follows.  In section II we discuss the theory and establish the relation between the cosmological constant and the CP violating term in general relativity.  In section III we show that there are physical effects from the CP violating term in GR if massive fermions are included. In section IV we combine the results of the previous sections to develop a quantum gravitational version of the Peccei Quinn mechanism to relax the cosmological constant.  Finally in section V we conclude with some future directions for calculating the chiral condensate expectation value in quantum gravity.
\section{The Theory}
General Relativity can be formulated in terms of a constrained topological field theory (TQFT).  In this paper we shall study a BF-theory for an $SO(5)$ gauge group.  BF theory is a topological field theory defined by the wedge product between a curvature two form $F(A)^{IJ}$ and an $SO(5)$ valued two form $B^{IJ}$ \cite{gary}.  When the symmetry breaks to $SO(4)$ the authors \cite{Smolina,laurent} showed, that the BF-theory becomes equivalent to general relativity with certain topological invariants.    While these topological terms do not affect the Einstein equations of motion classically, we will argue that a quantum mechanical effect can relax the cosmological constant in a manner similar to the Peccei-Quinn mechanism\cite{Peccei:1977hh}.  

Let $T^{IJ} = -T^{JI}$ be ten generators of the $SO(5)$ Lie algebra, where $I,J=1, ...,5$.  The basic dynamical variables are the $SO(5)$ connection $A^{IJ}$ and the $SO(5)$ valued 2-form $B^{IJ}$.  A two-form carries a pair of anti-symmetric internal indices $IJ$, with each index taking values from 0 to 4 and the lower case indices $i,j$ run from $i,j =1...4$. 

The theory we will study is gravity coupled to massive fermions and a pseudoscalar field.  Let us write the theory down and motivate each part afterwards:
\be \label{first} S_{tot} = S_{G} + S_{D} + S_{\phi} \ee
\begin{eqnarray} \label{gravityaction} S_{G} = -\int_{\cal{M}\rm} (B^{IJ}\wedge F_{IJ}- \frac{\beta}{2}B^{IJ} \wedge B_{IJ}
\nonumber\\
 - \frac{\alpha}{4}B_{IJ}\wedge B_{KL}\epsilon^{IJKL5}). \end{eqnarray}

\be S_{D}= \int_{M} e(\bar{\psi}_{1L} \tau^{A}E_{A}D\psi_{1L} +\bar{\psi}_{2L}\tau^{A}E_{A}D\psi_{2L}) \ee
and
\begin{eqnarray} \label{axionaction}
S_{\phi}  =\label{pseudo} \int_{M} -\frac{1}{2} \partial_{\mu}\phi\partial^{\mu}\phi
 + (\frac{\phi}{M})R_{ij}\wedge R^{ij} 
\nonumber\\
-if\partial_{\mu} \phi\bar{\psi}^{1}_{L}\gamma^{\mu}\psi^{1}_{L}
 - if\partial_{\mu} \phi \bar{\psi}^{2}_{L}\gamma^{\mu}\psi^{2}_{L} \end{eqnarray}
where $E_{A}=E^{\mu}\partial_{\mu}$ are the inverse vierbein vector fields, $f$ is a dimensionless constant and $F(A)^{IJ}$ the curvature 2-form
\be F^{IJ} = dA^{IJ} + A^{I}_{K}\wedge A^{K}_{J} \ee

Let us motivate each part of the action and show how they conspire to relax the cosmological constant non-perturbatively.  $S_{G}$ will be discussed in this section, $S_{D}$ in section III and $S_{\phi}$ in section IV.

The action $S_{G}$ in (\ref{gravityaction}) is equivalent to general relativity with a non-zero cosmological constant via a Macdowell-Mansouri mechanism where topological $F\wedge F$  theory undergoes a symmetry breaking from $SO(5)$ to $SO(4)$.  The case considered here arrives to general relativity from a $SO(5)$ $B\wedge F$ theory\cite{Smolina,laurent}.   
The authors of \cite{laurent} showed that the $B\wedge F$ theory spontaneously broken from 
$SO(5)$ to $SO(4)$ is equivalent to general relativity by introducing the $SO(4)$ connection $w^{ij} =A^{ij}$ and its curvature $R^{ij}=dw^{ij} + w^{i}_{k} \wedge w^{kj}.$  They also introduce a frame field $e^{i}=lA^{i5}$, where $l$ is a constant of dimension length, giving rise to a four-dimensional metric $g_{\mu\nu}=e^{i}_{\mu}e_{\nu j}.$
This gives the decomposition of the $SO(5)$ curvature:
  \be R^{ij}(w)= F^{ij}(A) - \frac{1}{l^{2}}e^{i}\wedge e^{j} \ee
In terms of this decomposition we can rewrite (\ref{gravityaction}) as:
\begin{eqnarray}
 S_{G}=  \label{gravityaction2} S_{P} + \int \frac{\alpha}{4(\alpha^{2} - \beta^{2})} R^{ij}(w) \wedge R^{kl}(w)\epsilon_{ijkl} 
\nonumber\\
- \frac{\beta}{2(\alpha^{2} -\beta^{2})} R^{ij}(w) \wedge R_{ij} (w) + \frac{C}{\beta}  
\end{eqnarray}
where $S_{p}$ is the Palatini action of general relativity, as will be discussed below and $C$ is the Nieh-Yan class \footnote{In eq (\ref{nieh}), $d_{w}e^{i} = de^{i}-w^{ij}e_{j}$.}
\be \label{nieh} C=d_{w}e^{i}\wedge d_{w}e_{i}- R^{ij}\wedge e_{i} \wedge e_{j}.\ee  
Notice that the parameter $\alpha$ in eq (\ref{gravityaction2}) corresponds to the magnitude of a fixed $SO(5)$ vector $v^{A}$.
\be \label{vec} v^{A} = (0,0,0,0,\alpha/2) \ee
We shall identify $\alpha$ with a pseudoscalar field $\phi$.  This field parameterizes the chiral rotations of the $SO(5)$ group.

\begin{eqnarray} \label{palatini}
S_{P} = -\frac{1}{2G}\int \large( R^{ij}(w)\wedge e^{k} \wedge e^{l}\epsilon_{ijkl} 
\nonumber\\
-\Lambda e^{i}\wedge e^{j} \wedge e^{k} \wedge e^{l} \epsilon_{ijkl} - \frac{2}{\gamma}R^{ij}(w) \wedge e_{i} \wedge e_{j} \large) 
\end{eqnarray} 
 is the Palatini action of general relativity with a nonzero cosmological constant and a nonzero Immirizi parameter $\gamma$, which is dimensionless.    The initial parameters $\alpha, \beta$ defined in eqs (\ref{vec}) and (\ref{first}), respectively, are related to the physical parameters as follows
\be 
\alpha=\frac{G\Lambda}{3(1-\gamma^{2})} ~~~~~ \beta=\frac{\gamma G\Lambda}{3(1-\gamma^{2})} .
 \ee

Let us focus on the the other terms of the action.  Crucial for our mechanism is the second term in eq (\ref{gravityaction2}), the topological CP violating term.  Using $\gamma$, we can rewrite this term as:  
\be \label{gsquare} S_{CP} =-\frac{3\gamma}{G\Lambda}\int_{M} R\wedge R 
.\ee

   This term is analogous to the CP violating term,$\tilde{\Theta} F\tilde{F}$, in QCD.
It has been proposed by Ashtekar, Balachandran and Jo, that there will be an ambiguous Yang-Mills instanton $\tilde{\Theta}$ angle term in the first order formulation because of large $SO(3,C)$ gauge transformations \cite{Ashtekar}.  The additional term in the action was suggested to be 

\be \label{rdual} 
S_{CPA}= -\frac{\Theta}{384\pi^{2}}\int R \wedge R\ee
Notice that the above action is identical to the second term in \ref{gsquare} if the $\Theta$ parameter scales inversely proportional to the cosmological constant, i.e if one identifies $\Theta$ in (\ref{rdual}) according to
\be  \label{Theta} \frac{\Theta}{384\pi^{2}} =\frac{3\gamma}{G\Lambda}. \ee

Now that we have discussed the connection between the gravitational $\Theta$ parameter and the cosmological constant we are ready to discuss the physical relevance of this connection.  As stated before, the $\Theta$ sector in quantum gravity corresponds to a one parameter family of vacuum states.   However, at this level $\Theta$ is fixed.  To allow $\Theta$ to be altered and dynamical, we introduce a set of chiral fermions.
\section{Chiral Rotations and Gravitational Axion}

Consider now the covariant coupling of gravity to chiral fermions.   Recall from equation (\ref{first}), 
\be S_{D} = -\frac{i}{2}\int_{M} e\bar{\Psi}\gamma^{A}E_{A}D\Psi + h.c. \ee
where the covariant derivative with respect to the spin connection, $w$, is 
\be D\Psi =dx^{\mu}(\partial_{\mu} + \frac{1}{2}w_{\mu BC}\cal{S}\rm^{BC})\Psi \ee
with the generators $\cal{S}\rm^{AB}=\frac{1}{4}[\gamma^{A},\gamma^{B}].$

We can work in a chiral basis by expressing the Dirac bispinor $\Psi$ in terms of two-component left and right handed Weyl spinors $\psi_{L,R}$. The chiral Dirac action which couples to the Palatini action now becomes
\be S_{D}= -i\int e(\bar{\psi}_{1L}\tau^{A}E_{A}D\psi_{1L} +\bar{\psi}_{2L}\tau^{A}E_{A}D\psi_{2L}) \ee
where
\be \psi_{R}=-i\tau^{2}\psi^{*}_{2L}. \ee

To understand the physical effects of the $\Theta$ term, consider the effect on the redefinition of all the fermion fields coupled to gravity.
\be \psi_{f} \rightarrow exp(i\gamma_{5}\alpha_{f})\psi_{f}, \ee
where $f$ is a flavor index and $\alpha_{f}$ are a set of real phases.   The effect on the measure of for path integrals over fermion fields is \cite{Chang}

\be [d\psi][d\bar{\psi}] \rightarrow exp\large(\frac{-i}{384\pi^{2}}\int R^{ij}\wedge R_{ij} \sum_{f} \alpha_{f} \large) [d\psi][d\bar{\psi}] \ee
Comparing this with eq (\ref{rdual}) we see that this phase shift in the fermion fields is equivalent to  shifting $\Theta$ by
\be \Theta \rightarrow \Theta + 2\sum_{f} \alpha_{f} . \ee
This phase rotation of the fermion fields also changes the mass terms in the Lagrangian

\be \cal{L}\rm_{m} = -\frac{1}{2}\sum_{f} \cal{M}\rm_{f} \bar{\psi}_{f,L}\psi_{f,L} - \frac{1}{2} \sum_{f}\cal{M}*\rm_{f}\bar{\psi}\rm_{f,R}\psi_{f,R} \ee

The phase redefinition changes the mass as follows
\be \cal{M}\rm_{f} \rightarrow exp(2i\alpha_{f})\cal{M}\rm_{f} \ee

It is important for our purposes to note that changing the path integration variables will not have any  physical effect, so observable quantities will only depend on the combination of $\Theta$ and the mass parameters $\cal{M}\rm_{f}$ 
\be exp(-i\Theta)\Pi_{f} \cal{M}\rm_{f} \ee 
restoring the dependence on $\gamma$ and $\Lambda$ this becomes.
\be \label{phase} exp(-i\frac{
\gamma 1154\pi^{2}}{2G\Lambda})\Pi_{f} \cal{M}\rm_{f} \ee 
From this expression and eq (\ref{Theta}) we see that the gravitational CP violating parameter, $\Theta$, is proportional to the ratio of the Immirzi parameter and the cosmological constant.  In the case of QCD, if the quark masses were to vanish, the Theta angle would have no effect and there would be no CP non-conservation Yang-Mills.  This observation will also hold in the gravitational case, hence we must have massive fermions coupling to gravity to implement our mechanism.

It is well known that if we define quark fields in QCD so that all $\cal{M}\rm_{f}$ are real, then a non-zero theta angle would produce a P and T non-conserving neutron electric  dipole moment proportional to $|\Theta|$ and $m_{\pi} ^{2}$ of the order
\be d_{n} \simeq |\Theta|em_{\pi}^{2}/m_{N}^{3} \simeq 10^{-24} e cm. \ee
In order to explain in a natural way why $\Theta$ is so small, Peccei and Quinn proposed that a pseudo-scalar field could dynamically relax to a minimum of the effective potential at which CP and P could be conserved.   Later Weinberg and Wilczek noted that it would require the existence of a light spinless particle, the axion.  The basic idea behind the axion theory is that there is a $U(1)$ symmetry which is spontaneously broken at energies much higher than those associated with QCD and is also broken by the anomaly involving the gluon fields.  In our context we want to propose a similar mechanism whose effective theory relaxes the gravitational $\Theta$ angle and therefore the cosmological constant by the presence of gravitational instantons.

\section{Quantum Gravitational PQ mechanism}

We shall now apply the Peccei-Quinn mechanism to general relativity.   Let us first briefly outline the mechanism.  We begin by introducing a pseudoscalar field in the action $S_{\phi}$ (\ref{pseudo}).  This field will play the role of the axion in QCD and will couple with two massive left handed Weyl fermions.    This pseudoscalar field arises from breaking an $SO(5)$ topological BF theory to $SO(4)$ as discussed in Section II.  According to arguments in the previous section, a shift in the mass term introduces a CP violating gravitational $\Theta$ parameter which is inversely proportional to the cosmological constant.  By generating a vev for the fermions, a potential is acquired for the pseudoscalar which relaxes the cosmological constant.    

Recall from Section III, gravity couples to left handed fermions via.
\be S_{D} = -\frac{i}{2}\int e(\bar{\psi} \tau^{A}E_{A}D\psi + h.c.) \ee
where 
\be D=d-iA_{a}\frac{\tau^{a}}{2} \ee
writing this terms of Weyl spinors we obtain
\be S_{D}= -i\int e(\psi^{\dagger}_{1L}\tau^{A}E_{A}D\psi_{1L} + \psi^{\dagger}_{2L}\tau^{A}E_{A}D\psi_{L2} ) \ee
where we have rewritten right handed spinors in terms of left handed ones.
\be \psi_{R}=-i\tau^{2}\psi^{*}_{2L} 
\ee
These chiral fermions couple covariantly to the pseudoscalar  field through the following aciton.       \begin{eqnarray} \label{axionaction}
 \cal{L}\rm_{\phi} =-\frac{1}{2} \partial_{\mu}\phi\partial^{\mu}\phi + [\frac{\phi}{M} + \frac{
K\gamma}{2G\Lambda}]R^{ij}\wedge R_{ij}
\nonumber\\
-i\partial_{\mu} \phi \bar{\psi}^{1}_{L}\gamma^{\mu}\psi^{1}_{L} - i\partial_{\mu} \phi \bar{\psi}^{2}_{L}\gamma^{\mu}\psi^{2}_{L} + \cal{L}\rm_{m}. 
\end{eqnarray}
where $K=1154\pi^{2}$ as in (\ref{phase}).

As stated in the last section we can eliminate the $\Theta$ term in eq(\ref{gravityaction2}) by redefining the mass term in the fermion Lagrangian density to 
\be \label{mass}  \cal{L}\rm _{m} = m\bar{\psi_{1L}}e^{-i\frac{f}{M}(\frac{M
K\gamma}{2G\Lambda} + \phi)\gamma_{5})}\psi_{1L} + m\bar{\psi_{2L}}e^{i\frac{f}{M}((\frac{M
K\gamma}{2G\Lambda}  + \phi)\gamma_{5})}\psi_{2L}  \ee 
where m is the fermion mass.
Now if the fermions condense and get a vev at some scale $M$
\be  <\bar{\psi^{1}_{L}}\psi^{1}_{L}>=<\bar{\psi^{2}_{L}}\psi^{2}_{L}>=\Delta \ee
where $\Delta \sim M^{3}.$
Then equation (\ref{mass}) becomes
\be  \label{potential} \cal{L}\rm_{m} =- \frac{1}{2} \partial_{\mu}\phi \partial^{\mu} \phi + m\Delta cos(\frac{f \phi'}{M}) \ee
where 
\be \phi'= <\phi> + \frac{M
K\gamma}{2G\Lambda}  \ee
Because the pseudoscalar Lagrangian has a shift symmetry $\phi' \rightarrow \phi' + \alpha$ we can redefine $\phi'$  at the minimum of the potential to yield the condition on the  cosmological constant in terms of the pseudoscalar field and the symmetry breaking scale.
Therefore, we can dynamically relax the cosmological constant.  At its minimum $<\phi>$ take the value.  
\be \label{relax} <\phi> \sim \frac{K\gamma M}{G\Lambda} .\ee

The above condition is not sufficient to cancel the bare cosmological constant, since according to eq(\ref{potential}) the height of the potential has to balance the difference between the bare and effective cosmological constant.  Hence, it allows the bare value of the cosmological constant to be shifted according to 
\be \label{min} \Lambda_{eff}=\Lambda + m \Delta cos(\phi') .\ee
When the fermionic chiral condensate acquires a vev, a potential for the pseudoscalar is generated.   Subsequently, the pseudoscalar minimizes its potential and gets a vev yielding $\Delta \ m cos\ (\phi') = -\Delta m$ for the potential.  Therefore when the potential of the pseudoscalar is minimized, eq(\ref{min}) becomes

\be \label{relation2} \Lambda_{eff} - \Lambda = \Delta m \ee 

We view this as a consistency relation for quantum gravity to relax the cosmological constant.  Hence, the theory should provide the value of the bare cosmological constant and the dynamics giving rise to the value of the gap $\Delta$ which cancels the bare cosmological constant.  With these two relations  (\ref{relax}) and (\ref {relation2}) we have a picture of how the cosmological constant can be dynamically suppressed in quantum gravity:

In order to solve the cosmological constant problem, the dynamics of the ground state of quantum gravity must self tune the vev of the gravitational fermionic condensate $\Delta$ to balance the magnitude of the bare cosmological constant.   This dynamical tuning must manifest itself  in the amplitude of the potential energy which is deep enough to cancel the bare cosmological constant when the pseudoscalar falls to its minimum.  In other words, this vev signify the formation of the energy gap of the fermion condensate whose mass 'eats' up the bare cosmological constant according to eq (\ref{relation2}).  It is interesting that the physics which sets the scale of the symmetry breaking (ie the condensation of the fermion bilinear) and the dynamics of the pseudoscalar field mutually work together to relax the cosmological constant.  
\section{Discussion}
 
We have presented a non-perturbative dynamical mechanism which relaxes the cosmological constant.  By realizing the connection between the cosmological constant, the gravitational $\Theta$ parameter and the Immirzi parameter, we were able to implement some wisdom from QCD in solving the strong CP problem, namely the Peccei-Quinn mechanism.  The solution of the cosmological constant can now be reformulated in-terms of a hierarchy between the mass of the gravitational axion and value of the chiral condensate energy gap $\Delta$.    

In light of this mechanism we can answer the question: Why does not the cosmological constant gravitate?  Simply put, there is a direct non-perturbative channel for a huge vacuum energy to transmute itself into a condensate. In doing so an energy gap opens up which offsets the value of the cosmological constant to a vanishing value and with respect to this new ground state the cosmological constant does not gravitate.  

For this mechanism to work, one needs to calculate the chiral condensate gap $\Delta$ in quantum gravity.  We propose that this calculation can be carried out using the Kodama State $\Psi(A)_K$ which corresponds to quantum gravity with a positive cosmological constant as well as de Sitter space, semiclassically\cite{Kodama:1990sc,Smolin}.  We conjecture that a modified Kodama state which includes the gravitational axion is the basis to evaluate the chiral condensate
\be <\bar{\psi}\gamma_{5}\psi>= <\Psi(A,\phi)_{K}|\bar{\psi}\gamma_{5}\psi|\Psi(A,\phi)_{K}> \ee
Recently, a modified normalizable Kodama State that describes Inflation was found, which included a scalar field \cite{Alexander2}.  We leave it up to a future investigations to see if this state is a candidate to evaluate the gap and hence see if the consistency relation (\ref{relation2}) is satisfied.

We would like to end with a physical speculation motivated by \cite{Alexander} which we believe might give insight into why quantum gravity can relax the
cosmological constant.  The authors of \cite{Alexander} argued that at  UV scales Minkowski space-time is unstable with respect to perturbative graviton exchange.  This statement is similar to the instability of the free fermi surface in superconductivity due to weak phonon exchange; and it is now understood that non-perturbative physics takes over and generates Cooper pairs from these exchanges between fermions while the system is driven to a true ground state that is purely non-perturbative.  In our case we can think of the vev of the fermion bilinears which couple to the gravitational field as Cooper pairs which signify the breakdown of a perturbative definition of the cosmological constant.  It is intriguing to know to what extent this analogy can be applied to Spin Networks.

\section{Acknowledgements}
It is a pleasure to thank Abhay Ashtekar, BJ Bjorken, Helen Quinn, Mohammad Sheikh-Jabbari and Marvin Weinstein for discussions.  I would especially like to thank Michael Peskin and Lee Smolin for looking over a few drafts of this paper and for challenging questions.  Even though the author has never met him in person, he would like to thank Chopin Soo for writing the influential series of papers on CTP properties of chiral gravity which impacted this work greatly.  Finally, I dedicate this work to my late grandfathers Stephen Belfon and Mustafa Mohammed-Ali.  I am grateful to the  US DOE for financially supporting this work under grant DE-AC03-76SF00515.

\end{document}